\documentclass[conference]{MIPRO}
\IEEEoverridecommandlockouts
\usepackage{cite}
\usepackage{amsmath,amssymb,amsfonts}
\usepackage{algorithmic}
\usepackage{graphicx}
\usepackage{textcomp}
\usepackage{xcolor}
\usepackage{booktabs}
\usepackage{todonotes}
\usepackage{subcaption}
\usepackage{balance}
\usepackage{url}

\begin{document}

\title{Scalability Evaluation of HPC Multi-GPU Training for ECG-based LLMs}

\author{\IEEEauthorblockN{Dimitar Mileski, Nikola Petrovski, Marjan Gusev}
\IEEEauthorblockA{\textit{Ss. Cyril and Methodius University in Skopje}, 
\textit{Faculty of Computer Science and Engineering} 1000, Skopje, North Macedonia   \\
\{dimitar.mileski, 
marjan.gushev\}@finki.ukim.mk} nikola.petrovski@students.finki.ukim.mk
}

\maketitle

\begin{abstract}
Training large language models requires extensive processing, made possible by many high-performance computing resources. This study compares multi-node and multi-GPU environments for training large language models of electrocardiograms. 
It provides a detailed mapping of current frameworks for distributed deep learning in multi-node and multi-GPU settings, including Horovod from Uber, DeepSpeed from Microsoft, and the built-in distributed capabilities of PyTorch and TensorFlow. We compare various multi-GPU setups for different dataset configurations, utilizing multiple HPC nodes independently and focusing on scalability, speedup, efficiency, and overhead. The analysis leverages HPC infrastructure with SLURM, Apptainer (Singularity) containers, CUDA, PyTorch, and shell scripts to support training workflows and automation.
We achieved a sub-linear speedup when scaling the number of GPUs, with values of 1.6x for two and 1.9x for four. 

\end{abstract}


\section{Introduction}
\label{sec-intro}

Finalizing the EU H2020 project entitled "AI Cardiologist – Alerting on Dangerous Arrhythmia" within the ELISE Open Call, we were challenged by the need to use extensive high performance computing (HPC) resources to build a Heart Language Model (HLM), as a transformer-based model \cite{tudjarski2025transformer}.
Our goal was to compare the execution of the training process through multi-GPU per node and multiple HPC nodes independently.
 A transformer-based foundation model was trained using electrocardiogram (ECG) data, tokenizing heartbeat sequences to construct a language representation of heart rhythms \cite{tudjarski2025transformer}. The model was fine-tuned using annotated benchmark databases and evaluated across various datasets to determine its ability to generalize to new data. 

This research continues the training of HLM through the EuroHPC Benchmark Access call from the EuroHPC Joint Undertaking, utilizing the Luxembourg national supercomputer, MeluXina, at LuxProvide. Instead of building a model on ECG annotations, it builds another HLM on ECG samples surrounding a heartbeat.
The HLM datasets include ECG benchmarks consisting of 50,067,387 heartbeats. Training multiple foundation models and fine-tuning different parameters requires multi-GPU and multi-node training to shorten the overall research and development time. Projects utilizing HPC resources are constrained by allocation periods and node-hour limits per month, requiring careful distribution of computing time across the project's duration. This is especially true in EuroHPC initiatives, where node hours must be evenly consumed throughout the project's lifecycle.

The structure of the paper is as follows: Section~\ref{sec-state} analyzes the similarities and differences in the state-of-the-art. Methods in Section~\ref{sec-methods} specify the solution architecture and workflow, experiments, and the evaluation metrics to evaluate the results. The results presented in Section~\ref{subsec-results} are evaluated and discussed in Section~\ref{sec-discussion}. Section~\ref{sec-conclusion} presents the conclusion and future work.

%
%

\section{State-of-the-art}
\label{sec-state}

As LLMs grow in size, in our case, the heart language model often reaches hundreds of billions of parameters with increased computational demands, requiring advanced hardware solutions for efficient training and inference. Multi-GPU configurations enable distributed computation within a single node, reducing training time through parallelism. Multi-instance GPU technology, such as Nvidia’s MIG, optimizes resource allocation by allowing multiple independent workloads to run on a single GPU, improving efficiency in inference-serving environments. For large-scale training, multi-node setups distribute workloads across multiple machines, utilizing high-speed interconnects to synchronize massive models efficiently.

%
\subsection{Multi-GPU: Nvidia}
NVIDIA's Nvidia Collective Communication Library (NCCL) optimizes multi-GPU and multi-node communication for deep learning (DL) integrated into major frameworks like PyTorch, TensorFlow, and MxNet to improve training efficiency \cite{weingram2023xccl}. 
The module torch.distributed in PyTorch enables efficient multi-GPU and multi-node training with three backends: NCCL (for GPUs), Gloo (for CPUs and fallback GPU support), and MPI (optional for CPU/GPU with InfiniBand). DistributedDataParallel (DDP) is part of torch.distributed for synchronous training, where each process maintains its optimizer to reduce synchronization overhead. 

Data loaders manage data ingestion, pre-processing, and delivery to the model during training. In distributed training (e.g., multi-GPU or multi-node setups), they ensure parallel data loading, synchronization, optimized I/O, and pre-processing to keep GPUs fully utilized and avoid idle time. They also use sharding to split the dataset into non-overlapping chunks, batching, and prefetching for each worker. DALI (NVIDIA Data Loading Library) is a high-performance data loading and augmentation library designed for DL frameworks like PyTorch and TensorFlow. 

Aach et al. \cite{aach2023large} compare distributed DL frameworks' performance and scalability, demonstrating that the NVIDIA DALI data loader significantly reduces training time. DALI, a high-performance data loader, outperforms traditional PyTorch data loaders by up to 3.4× in loading times for large datasets \cite{martinez2024high}. Using data loading tools such as binary data formats and NVIDIA DALI can improve the training time of deep neural networks by 20–40\% \cite{zolnouri2020importance}. 
The native data loader cannot match the performance of the DALI data loader analyzing three frameworks \cite{aach2023large}.

PTX (Parallel Thread Execution) is an intermediate assembly language for NVIDIA GPUs. It is used in CUDA programming to provide a low-level, human-readable representation of how threads execute on the GPU \cite{lustig2019formal}. 
DeepSeek \cite{liu2024deepseek} enhanced AI training efficiency by replacing CUDA with Nvidia’s low-level PTX programming for specific functions, enabling fine-tuned GPU optimizations at the register and thread level for improved performance. DeepSeek claims to have trained a competitive AI model using 11 times less GPU compute than major players like OpenAI and Meta \cite{liu2024deepseek}. Their DeepSeek-V3 Mixture-of-Experts (MoE) model, with 671 billion parameters, was trained on 2,048 Nvidia H800 GPUs in just two months \cite{liu2024deepseek}, significantly reducing costs and resource demands. DeepSeek-V3 is trained on a cluster equipped with 2048 NVIDIA H800 GPUs. Each node in the H800 cluster contains 8 GPUs connected by NVLink and NVSwitch within nodes. Across different nodes, InfiniBand (IB) interconnects are utilized to facilitate communications. DeepSeek auto-tunes the communication chunk size, effectively minimizing L2 cache usage and reducing interference with other Streaming Multiprocessors.

In this research, we chose CUDA because the Elise project HLM codebase is written in PyTorch. PyTorch provides built-in support for DDP and CUDA wrapper functions, allowing us to leverage GPU acceleration without writing PTX code. PyTorch internally uses NCCL for communication in distributed training. For data loading, we rely on the native PyTorch DataLoader.
 
%
\subsection{Multi Instance (Containers)-GPU: Nvidia} 

DL training is a resource-intensive process that heavily relies on GPUs, but not all models fully utilize the capabilities of modern, powerful GPUs. NVIDIA's Multi-Instance GPU (MIG) technology allows a single GPU to be partitioned, making it more efficient for workloads that don't need an entire GPU's full memory and computational power. NVIDIA's Multi-Instance GPU (MIG) technology allows partitioning a single GPU into multiple virtual GPUs, enabling simultaneous use by multiple users. While MIG provides isolation and security benefits, it can lead to performance degradation, with up to 54 times slower execution in some cases \cite{de2024avaliaccao}. However, for workloads that don't require complete GPU resources, MIG can significantly improve utilization and throughput \cite{Kaas2022DeepLT}. A compiler-assisted approach can further enhance MIG's efficiency by packing jobs into partitions when isolation isn't necessary, resulting in 1.45x throughput improvement \cite{porter2022compiler}. MIG can offer up to 3 times the throughput for training small DL models but provides only marginal benefits for medium and large workloads \cite{Kaas2022DeepLT}. Optimizing MIG configurations for serving Deep Neural Networks is challenging, but algorithms like MIG-serving can save up to 40\% of GPUs while maintaining throughput \cite{tan2021serving}.

In this research, we do not use Multi-Instance GPU, as NVDashboard \cite{nvidiaDashboardsJupyter} indicates good GPU utilization, and some of the referenced works suggest that MIG may degrade performance.

%
\subsection{Multi-Node: Distributed DL}
Horovod \cite{sergeev2018horovod} is an open-source distributed DL training framework designed to simplify scaling training across multiple GPUs and nodes, leveraging the Ring-AllReduce algorithm to synchronize gradients across different workers. 
Horovod seamlessly integrates with popular DL frameworks like TensorFlow, PyTorch, and Apache MXNet, enabling users to achieve substantial performance gains with minimal code changes. It also facilitates distributed training for large models by supporting both data and model parallelism.

DeepSpeed \cite{rasley2020deepspeed} is a DL optimization library developed by Microsoft that significantly enhances the training of large-scale models. It enables training models with billions of parameters, such as BERT-large, in under an hour using 1,024 GPUs, while conventional methods struggle with smaller models. The library features the Zero Redundancy Optimizer, which allows models with up to 13 billion parameters to be trained on a single GPU, reducing memory usage by up to 8x compared to existing frameworks. 
DeepSpeed supports various parallelism techniques, such as 3D parallelism, to manage models with trillions of parameters, achieving throughput improvements of up to 10x. Additionally, its advanced optimizers, like 1-bit Adam, reduce communication volume by up to 26x \cite{rasley2020deepspeed}.

PyTorch Distributed enables parallel and distributed training across multiple nodes or GPUs. PyTorch provides a distributed data parallel module that enables near-linear scalability of model training using up to 256 GPUs \cite{li2020pytorch}. Apache MXNet offers high flexibility with automatic GPU parallelization, making it well-suited for research and production. Deeplearning4j, a Java-based framework, is tailored for Java developers and integrates well with significant data ecosystems. Elephas integrates Keras with Apache Spark, leveraging Spark's data parallelism for distributed training. HAI-LLM is optimized for large-scale language model training, focusing on efficient multi-GPU and multi-node setups. Frameworks like Caffe, Chainer, and Theano do not offer native distribution support.
In contrast, others like CNTK, DL4j, and PyTorch provide centralized and decentralized training, with varying levels of support for synchronous and asynchronous training \cite{mayer2020scalable}. Some frameworks, such as MXNet and TensorFlow, support model quantization, while others, like PyTorch and Keras, do not. Communication scheduling is largely unsupported across these frameworks, except for a few. The community support, measured through GitHub repositories and StackOverflow mentions, varies significantly, with TensorFlow having the most significant presence, followed by PyTorch \cite{mayer2020scalable}.

For multi-GPU training, we use CUDA, NCCL, PyTorch-integrated CUDA functions, and NVDashboard \cite{nvidiaDashboardsJupyter}. We do not use PTX to optimize parallel threads with assembly language, nor do we utilize MIG for multi-instance GPU with CUDA. While we tested PyTorch Distributed for multi-node training, its speedup and efficiency were suboptimal, as detailed in the results and future work section. We trained various foundation and fine-tuning models on different ECG datasets, running each node independently while utilizing all four Tesla A100 GPUs per node (a total of four HPC nodes simultaneously, with 16 Nvidia A100 GPUs).

%
%
\section{Methods}
\label{sec-methods}

This section outlines the methods used to evaluate the performance of foundational and fine-tuned model training of ECG-based large language models (LLMs) in different hardware configurations (CPUs and GPUs) within a single HPC node and multiple independent HPC nodes. First, we present the workflow architecture, detailing the key components. Then, we specify the experimental setup, including model configuration and dataset parameters. Finally, we define the evaluation methodology, focusing on training time, speedup, and efficiency metrics to assess scalability and performance improvements.

%
\subsection{Workflow architecture}
\label{subsec-architecture}

\begin{figure}[t]
    \centering
    \includegraphics[width=.8\columnwidth]{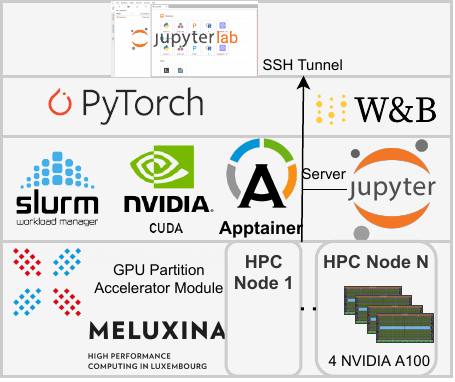}  
    \caption{System Architecture}
    \label{fig-Architecture}
\end{figure}

Figure~\ref{fig-Architecture} illustrates the system architecture, starting with the HPC hardware accelerator layer, which utilizes the Meluxina supercomputer hosted at LuxProvide. This system consists of multiple accelerator (GPU) nodes, each equipped with four A100-40 GPUs. Detailed specifications of the hardware layer are provided in Table~\ref{tab-meluxina}. We rely on the SLURM Workload Manager in the software layer, where we write bash scripts to run SLURM batch jobs. The HLM codebase is in Python and includes monitoring and tracking code snippets that send data to the WandB Cloud Platform. We use Apptainer (formerly Singularity) images with CUDA and PyTorch to ensure portability and compatibility with HPC environments. Additionally, JupyterLab is deployed on one of the HPC nodes, and researchers can access it remotely from their laptops through an SSH tunnel.
\begin{table}[!t]
\caption{MeluXina GPU Partition: 200 CPU-GPU Hybrid Nodes with 800 GPU-AI Accelerators}
\label{tab-meluxina}
\centering
\resizebox{\columnwidth}{!}{
\begin{tabular}{|l|l|}
\hline
\textbf{Component}               & \textbf{Specification}                 \\ \hline
\# of Nodes                  & 200 CPU-GPU hybrid nodes               \\ \hline
Total \# of GPUs              & 800 GPU-AI accelerators                \\ \hline
CPUs (2 AMD Rome)	& 32 cores@2.35 GHz, 128 HT cores total \\ \hline
GPUs                               & 4 NVIDIA A100-40 GPUs                  \\ \hline
RAM                                & 512 GB RAM                             \\ \hline
Storage                            & SSD 1.92 TB                            \\ \hline
Network                            & 2 HDRcards InfiniBand network          \\ \hline
\end{tabular}
}
\end{table}

\begin{figure}[t]
    \centering
    \includegraphics[width=.8\columnwidth]{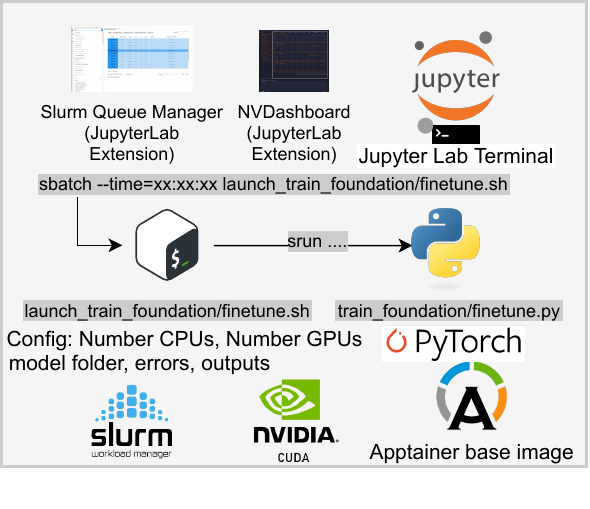}  
    \caption{Experiment Workflow}
    \label{fig-Workflow}
\end{figure}

An sbatch command from the JupyterLab terminal initiates the code execution (Figure~\ref{fig-Workflow}) with a workflow that follows a compilation and execution path. The process begins when the shell script (.sh) is submitted to the Slurm workload manager, which handles resource allocation and job scheduling. This shell script triggers the Python script (.py) containing the training code. The Python code, written using PyTorch, gets interpreted, and the PyTorch operations are compiled into CUDA kernels through NVIDIA's CUDA toolkit if GPU operations are involved. 
This entire process runs within an Apptainer container environment, which ensures all necessary dependencies and libraries are available and consistent.

A shell script invokes a wrapper Python script (Figure~\ref{fig-Workflow}) to start a SLURM batch job 
\begin{footnotesize}
\texttt{sbatch --time=xx:xx:xx launch\_train\_foundation.sh}.
\end{footnotesize}
We use an Apptainer base image with PyTorch and CUDA and specific Apptainer images built on top of it, including the Python libraries. SLURM batch jobs are submitted and monitored using the SLURM Queue Manager (JupyterLab extension) and NVDashboard (JupyterLab extension) for GPU usage monitoring. 

Researchers' SLURM batch jobs run using the Jupyter Terminal or an SSH client (Windows/Linux). Jupyter File Manager and the WinSCP client on Windows manage the files. Python scripts are written in JupyterLab with a Python kernel, and port forwarding is used to access JupyterLab on the MeluXina HPC from the researcher's laptop.

We encountered challenges setting up the workflow, configuring SLURM batch jobs, configuring up Apptainer images (Pytorch and CUDA), installing JupyterLab along with its plugins (NVDashboard and SLURM Queue Manager), and configuring port forwarding to JupyterLab on MeluXina. 
To address these challenges, we upgraded our prior knowledge of HPC and Cloud Systems with customized specifications in MeluXina's user guides and assistance from the MeluXina support team (LXP Service Desk).

Python scripts send data to the WANDB cloud platform, monitoring metrics and logs about the training process. We used NVDashboard to observe the utilization of GPUs and VizTracer for profiling to identify bottlenecks in the code.

\begin{figure*}[htbp]
    \centering
    \begin{subfigure}{0.32\textwidth}
        \centering
        \includegraphics[width=\linewidth]{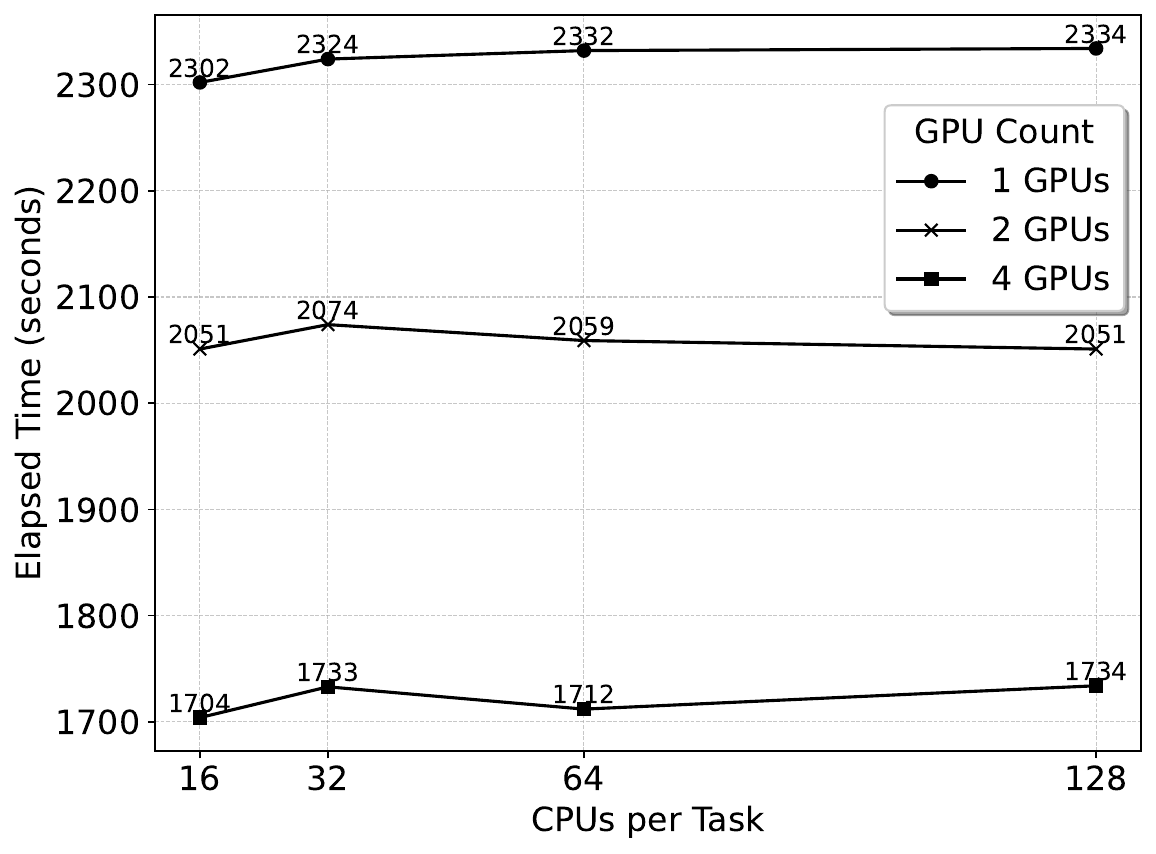}
        \caption{Foundation VBD MITDB 128 MELUXINA V40}
        \label{fig-sub1}
    \end{subfigure}
    \hfill
    \begin{subfigure}{0.32\textwidth}
        \centering
        \includegraphics[width=\linewidth]{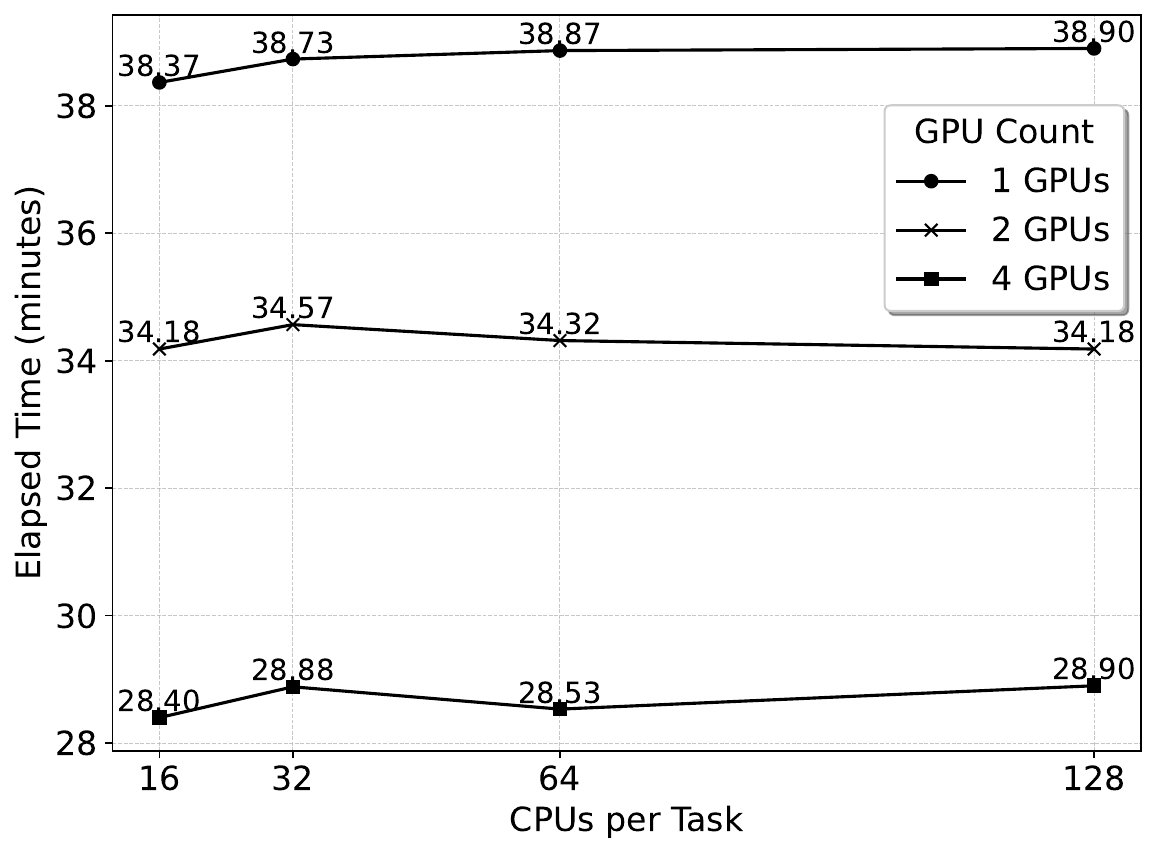}
        \caption{Foundation VBD MITDB 128 MELUXINA V40}
        \label{fig-sub2}
    \end{subfigure}
    \hfill
    \begin{subfigure}{0.32\textwidth}
        \centering
        \includegraphics[width=\linewidth]{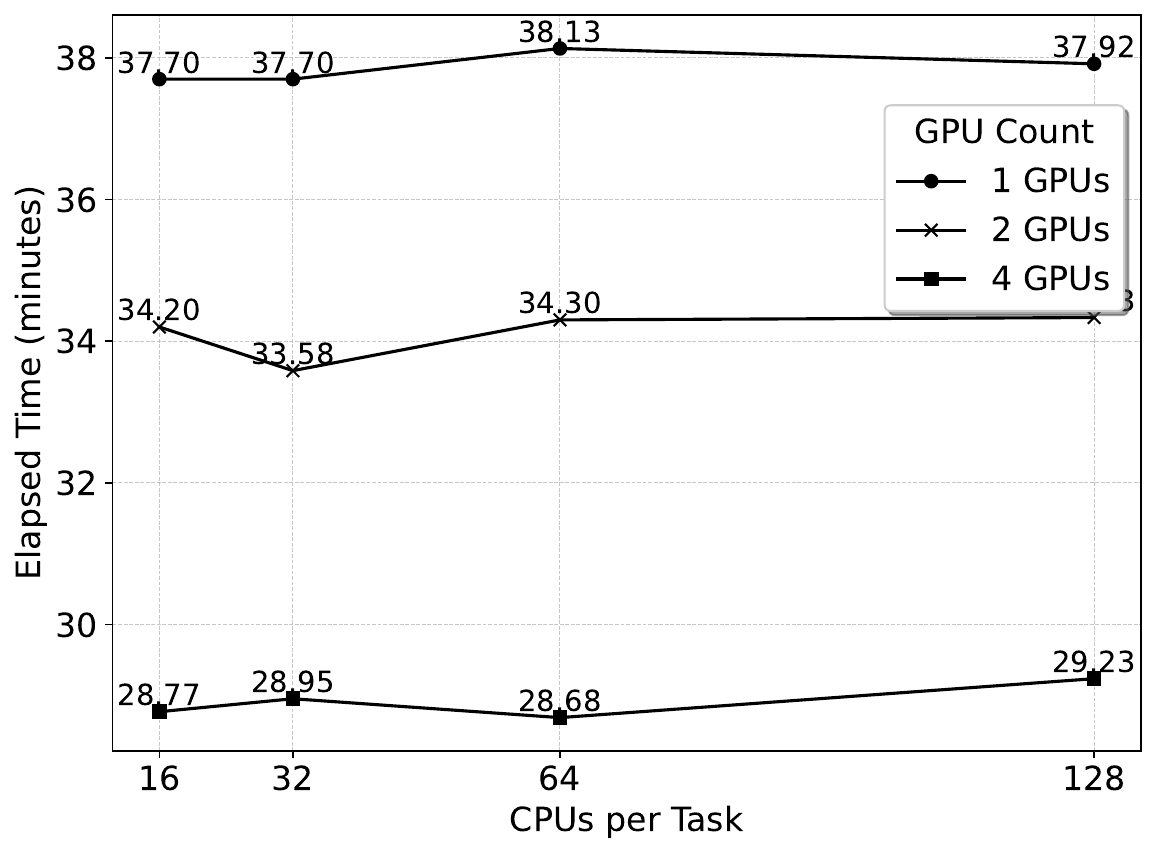}
        \caption{Foundation VBD MITDB 1024 MELUXINA V40}
        \label{fig-sub3}
    \end{subfigure}
    \caption{Comparison of training times across different CPU and GPU configurations.}
    \label{fig-wall-clock-time}
\end{figure*}

\begin{figure*}[htbp]
    \centering
    \begin{subfigure}{0.32\textwidth}
        \centering
        \includegraphics[width=\linewidth]{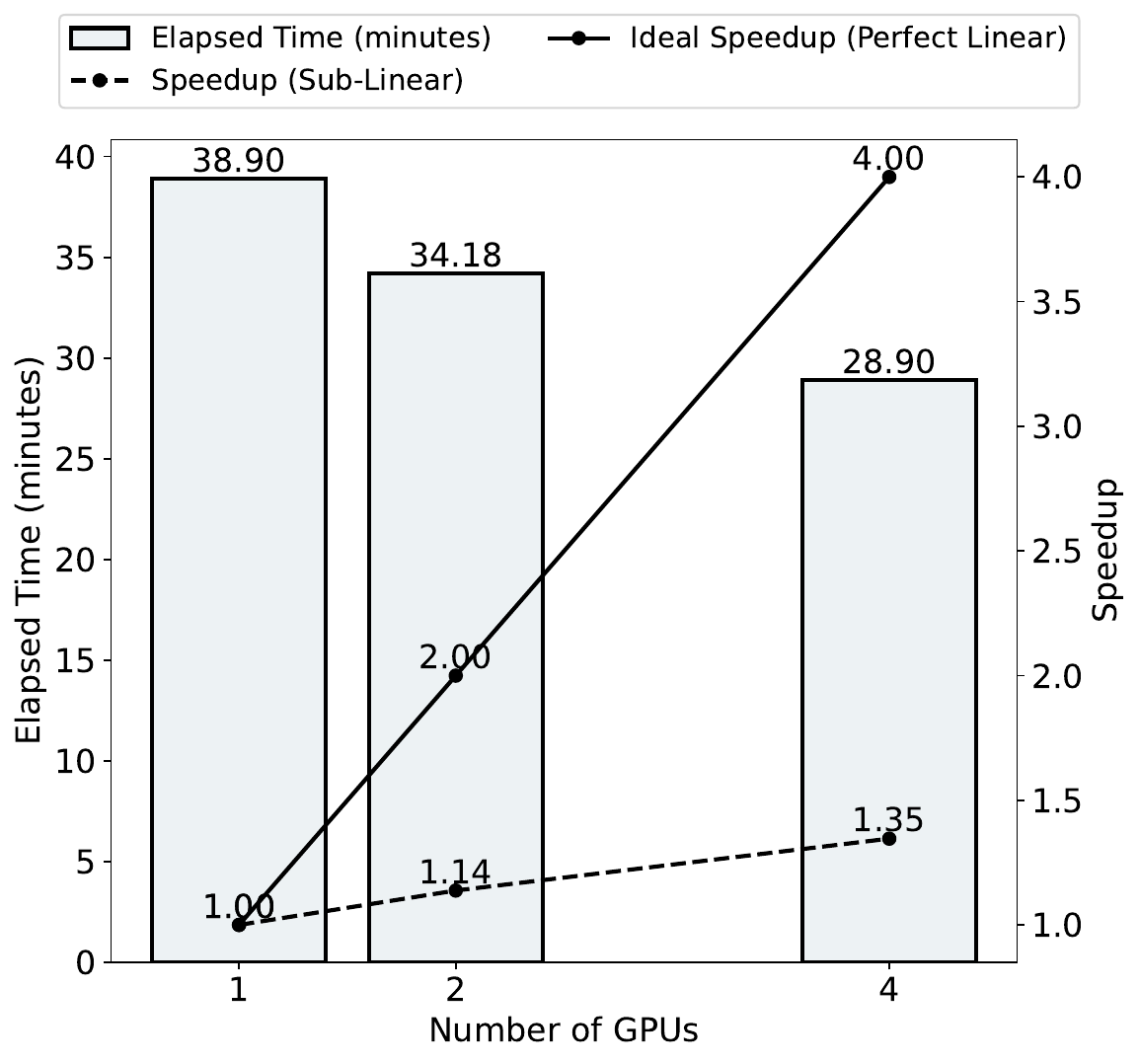}
        \caption{Foundation VBD MITDB 128 MELUXINA V40}
        \label{fig-gpu_sub1}
    \end{subfigure}
    \hfill
    \begin{subfigure}{0.32\textwidth}
        \centering
        \includegraphics[width=\linewidth]{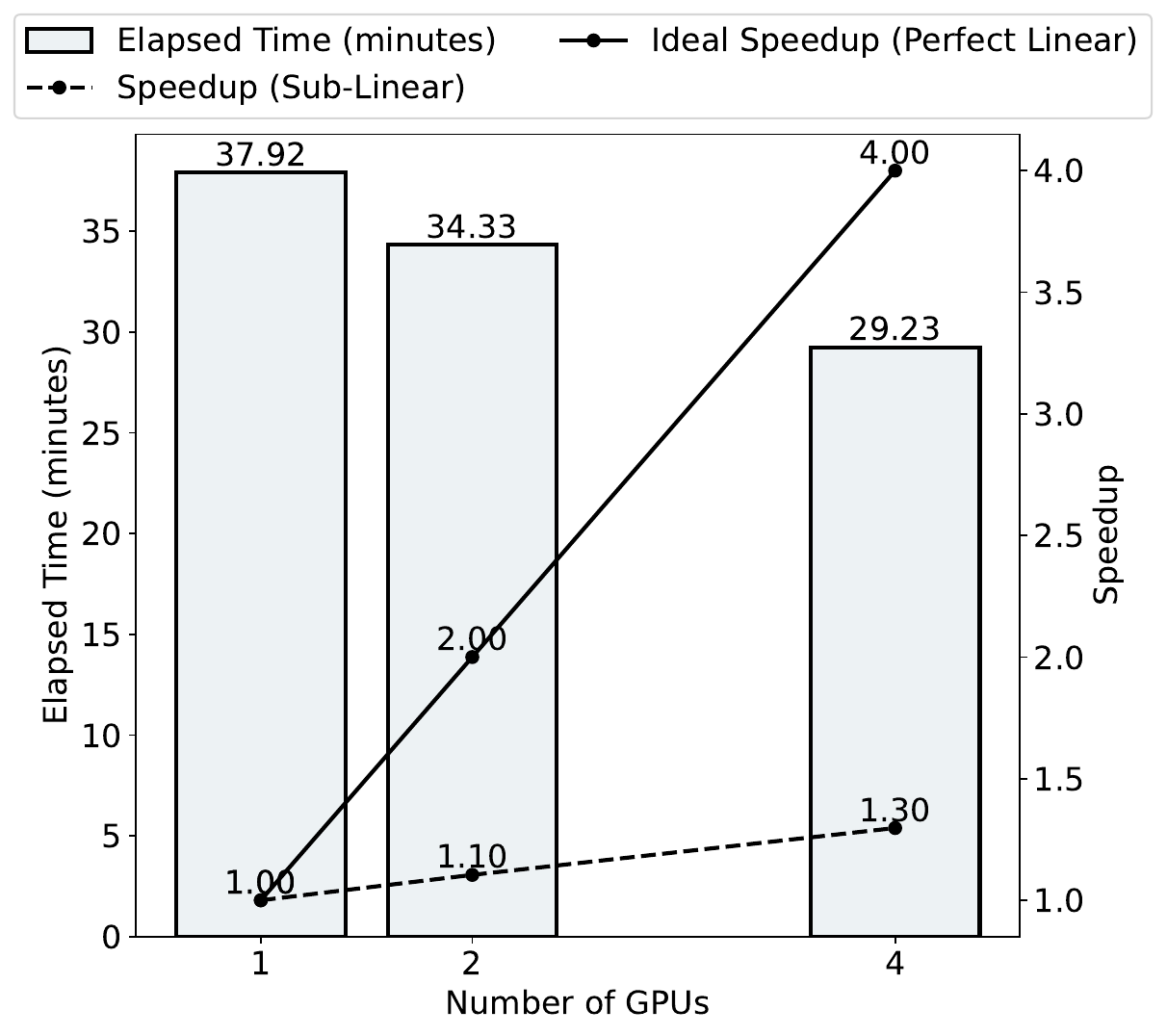}
        \caption{Foundation VBD MITDB 1024 MELUXINA V40}
        \label{fig-gpu_sub2}
    \end{subfigure}
    \hfill
    \begin{subfigure}{0.32\textwidth}
        \centering
        \includegraphics[width=\linewidth]{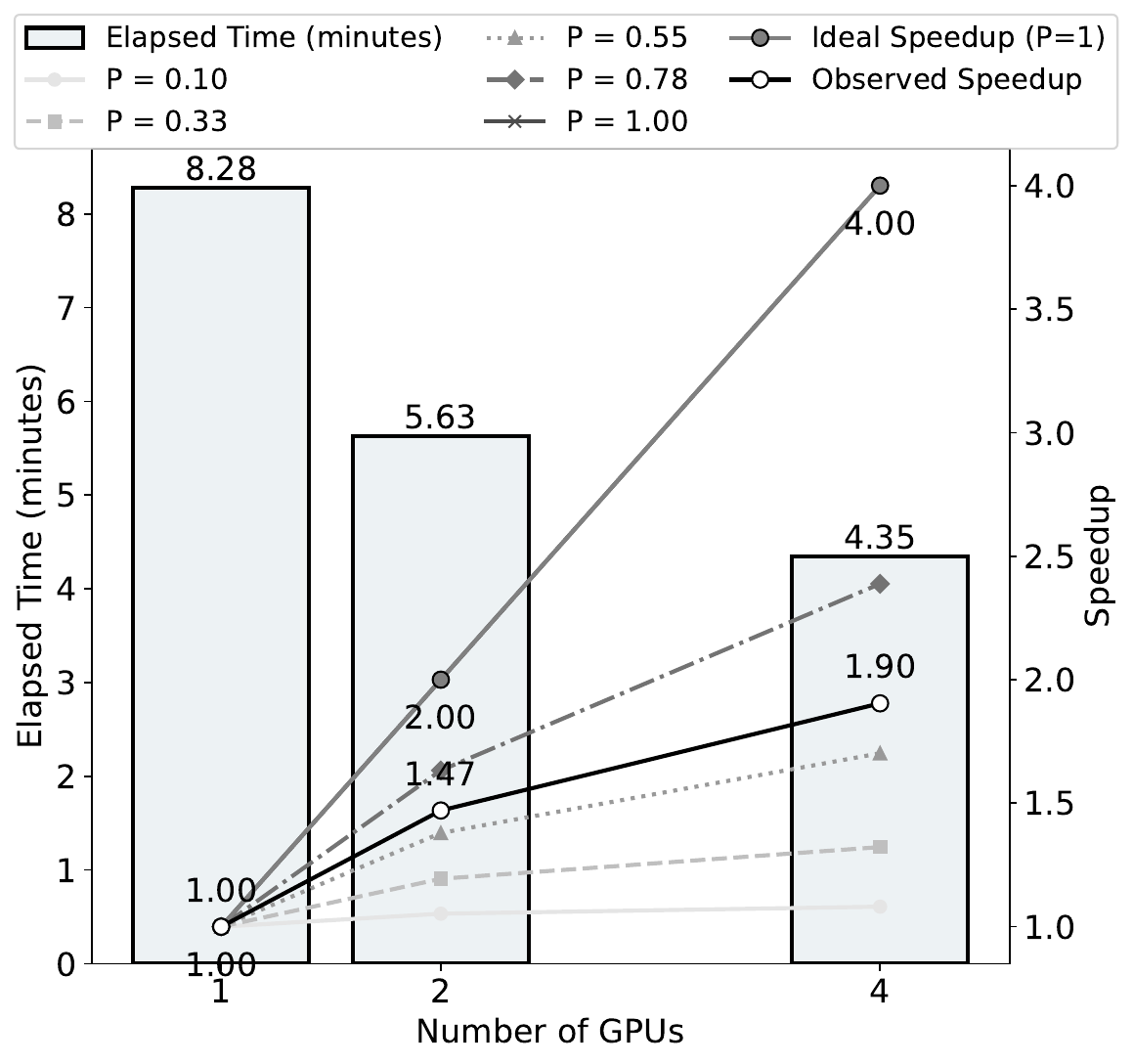}
        \caption{Finetune VBD MITDB 128 MELUXINA V40}
        \label{fig-gpu_sub3}
    \end{subfigure}
    \caption{Comparison of training time, speedup, and ideal speedup across different GPU configurations.}
    \label{fig-speedup}
\end{figure*}

\begin{figure}[htbp]
    \centering
    \includegraphics[width=\columnwidth]{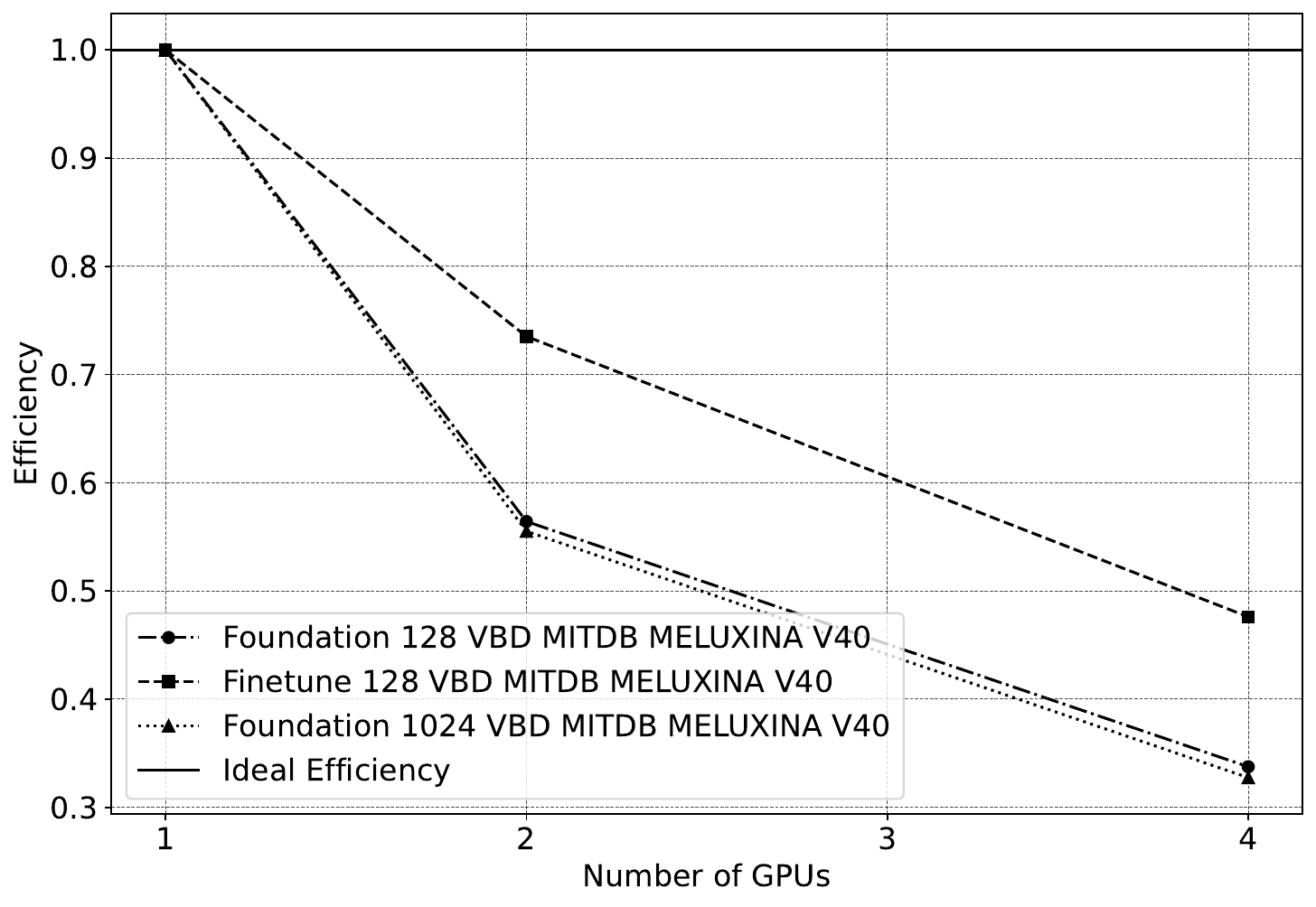}  
    \caption{Efficiency}
    \label{fig-efficiency}
\end{figure}

\begin{figure}[htbp]
    \centering
    \includegraphics[width=\columnwidth]{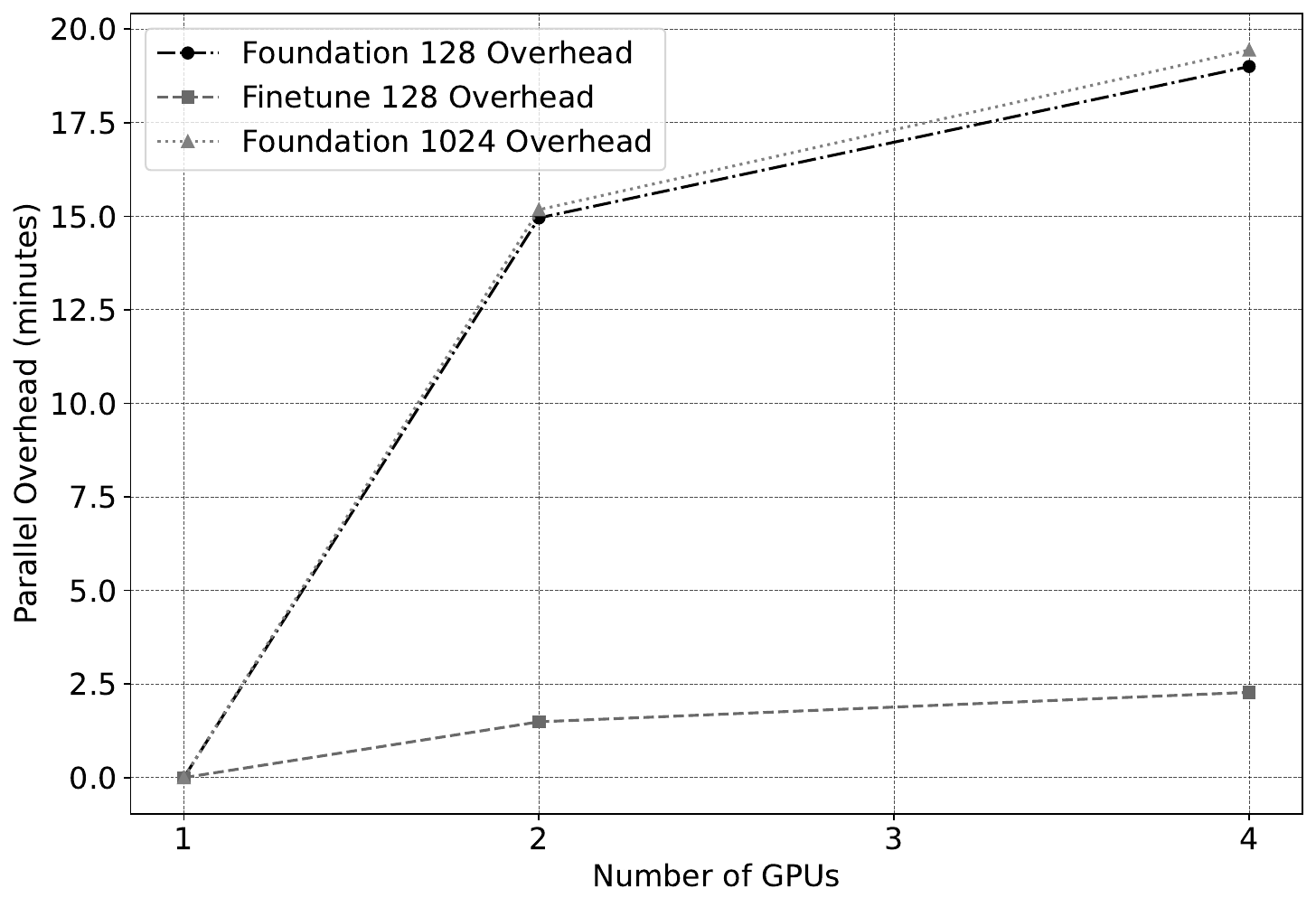}  
    \caption{Overhead}
    \label{fig-Overhead}
\end{figure}

%
\subsection{Experiments}
\label{subsec-experiments}

\begin{itemize}
  \item \textbf{Datasets:} 16 ECG datasets, totaling 272 GB.
  \item \textbf{Sliding Window Sizes:} We evaluate with various sliding window sizes (128 ... 1024).
  \item \textbf{Hardware Configurations:}
  \begin{itemize}
    \item \textbf{CPU Configurations:} 1, 2, 4, 16, 32, 64, and 128 CPU cores.
    \item \textbf{GPU Configurations:} 1, 2, and 4 GPUs.
  \end{itemize}
\end{itemize}

%
\subsection{Evaluation methodology}
\label{subsec-eval}

We evaluated the {\em speedup} as an improvement in training time compared to the baseline configuration (1 GPU) by calculating (\ref{eq-speedup}), where
$T_{\text{baseline}}$ is the training time using 1 GPU (1N1GPU), $T_{\text{test}}$ is the training time using 2 GPUs (1N2GPU) or 4 GPUs (1N4GPU) and $N_{\text{GPUs}}$ refers to the total number of GPUs used in the test configuration.
\begin{equation}
\label{eq-speedup}
\text{Speedup} = \frac{T_{\text{baseline}}}{T_{\text{test}}}
\end{equation}

The {\em efficiency} measures how effectively the additional GPUs improve performance, calculated by (\ref{eq-eff}). Efficiency = 1 means ideal scaling (each additional GPU contributes fully), lower values refer to reduced scaling due to communication overhead, memory bandwidth limits, or other inefficiencies, and higher values, although uncommon, can happen due to algorithmic optimizations that improve cache/memory access patterns.
\begin{equation}
\label{eq-eff}
\text{Efficiency} = \frac{\text{Speedup}}{\text{$N_{\text{GPUs}}$}}
\end{equation}

The {\em overhead} metric measures the extra time spent due to communication, synchronization, and inefficiencies, calculated by (\ref{eq-overhead}). A lower overhead is better, indicating better parallelization. If overhead increases significantly with more GPUs, it suggests inefficiencies in communication or workload distribution. Ideal scaling would show minimal overhead as GPUs increase.
\begin{equation}
\label{eq-overhead}
\text{Overhead} = T_{\text{test}} - \left(\frac{T_{\text{baseline}}}{N_{\text{GPUs}}}\right)
\end{equation}

%
%
\section{Results}
\label{subsec-results}

Figure~\ref{fig-wall-clock-time} compares training times across various CPU and GPU configurations. Each subfigure (a, b, and c) displays the relationship between the number of CPUs per task (x-axis) and the training time in seconds (y-axis). Different configurations are analyzed based on the number of GPUs utilized during training. The graphs highlight how training time varies when increasing the number of CPUs per task under different GPU settings.

Speedup charts in Figure~\ref{fig-speedup} illustrate the training time, speedup, and ideal speedup across different GPU configurations. The x-axis represents the number of GPUs used, while the left y-axis indicates the training time in seconds. Additionally, the right y-axis shows the speedup compared to the baseline (single GPU case). The bars depict the training time, whereas the speedup curves compare the observed speedup to the ideal scaling scenario, demonstrating the efficiency of parallelization.

The x-axis in Figure~\ref{fig-efficiency} represents the number of GPUs to evaluate the efficiency of different GPU configurations on the y-axis, defined as the achieved speedup relative to the ideal speedup. The graph includes multiple configurations, showing efficiency changes as more GPUs are added.

Figure~\ref{fig-Overhead} analyzes the overhead associated with different GPU configurations. The x-axis represents the number of GPUs, while the y-axis measures the relative computational overhead. The graph compares multiple configurations, showing how overhead increases with the number of GPUs, providing insights into the scaling limitations of the training process.

%
%
\section{Discussion}
\label{sec-discussion}

We ran multiple models with different configurations to develop foundation and fine-tuning models and presented the corresponding speedups across different numbers of CPUs and GPUs. Performance metrics include wall-clock time (execution time), speedup, efficiency, and overhead.

The first experiment is on a single dataset, varying the number of CPUs, GPUs, and sliding window sizes. Then, we execute the experiment on all remaining datasets using optimal values for CPUs and GPUs for 16 datasets with eight different model window widths. The largest dataset is 17 GB, and the total size of all datasets is 272 GB. It is worth mentioning that these datasets consist of ECG signals, and the 272 GB of ECG data is a significant value for the training process. Unlike text data, which can often be scraped from public sources, ECG data requires specialized collection and processing following the GDPR and HIPPA regulations.

We observed that for a single dataset, when varying the number of CPUs, GPUs, and sliding window sizes, the maximum speedup was achieved with a sliding window size of 128. These plots show that our workflow does not achieve speedup when increasing the CPU count beyond 16 (i.e., from 16 to 31, 64, or 128 CPUs). We observe sub-linear speedup across all sliding window sizes and datasets when increasing the GPU count from one to two and up to a maximum of four GPUs (as in the MeluXina GPU node partition). The highest speedup achieved was 1.9× for four GPUs on the fine-tuning model with a sliding window size 128.

Using our previous HLM codebase without considering minor code optimizations, we achieved a 2× speedup on four GPUs. While this speedup is sub-linear, it significantly reduces the time required to run multiple experiment configurations across different datasets. Without node-hour availability, many of these configurations would be infeasible, especially in the final month of allocation, when we run experiments extensively.

For example, in one experiment configuration with 11 out of 16 datasets, data (token) generation took 11 hours. Without achieving the 2× speedup on 4 GPUs, the same experiment would have taken nearly 22 hours, consuming valuable node hours that could have been used for other experiments. The best efficiency observed was 0.75 for 2 GPUs (where the ideal value is 1) and approximately 0.5 for 4 GPUs (ideal value: 1).

\section{Conclusion}
\label{sec-conclusion}

Our efforts to develop a generative AI-based heart monitoring solution using wearable ECG sensors for arrhythmia detection have yielded valuable insights into the challenges and opportunities, conducting experiments on the MeluXina HPC platform with multiple configurations of CPUs, GPUs, and sliding window sizes across 16 datasets. Through this process, we observed that the optimal sliding window size for maximum performance was 128, while the scalability of CPUs beyond 16 offered no significant speedup. In contrast, we achieved a sub-linear speedup with GPUs, where using two GPUs resulted in a 1.6× speedup, and scaling to four GPUs provided a maximum speedup of 1.9× (below the theoretical 4× speedup for perfect linear scaling). 

Despite the lower efficiency of 0.75 for two GPUs and 0.5 for four GPUs, our workflow effectively managed the large-scale computational requirements, and we successfully conducted experiments across 16 datasets with various model configurations. These results highlight the significant potential of HPC in accelerating the development of AI models for healthcare applications but also emphasize the need for further code optimization and GPU scaling improvements. As we move forward, we plan to leverage the experience gained from these experiments to refine our approach, applying for the EuroHPC Development Access Call to continue enhancing our generative AI heart monitoring solution, which we believe has the potential to revolutionize patient care, particularly in outpatient settings, while also driving business growth in the EU healthcare market.

The possible key optimizations for future work include improving the code and eliminating the bottleneck functions in PTX to achieve finer control over GPU execution and enhance performance. This will allow us to optimize low-level CUDA operations critical to our workload.
Additionally, we intend to integrate NVIDIA DALI for data preprocessing and loading. To maximize GPU resource utilization, we will explore Multi-Instance GPU (MIG), enabling efficient sharing of a single GPU across multiple processes. Furthermore, we plan to leverage PyTorch Distributed for multi-node training, ensuring efficient scalability across multiple nodes.

%
%
 \section*{Acknowledgment}
 \label{sec-ack}
We acknowledge EuroHPC Joint Undertaking for awarding us access to MeluXina at LuxProvide, Luxembourg, through an EuroHPC Benchmark Access call. The simulations were performed on the Luxembourg national supercomputer MeluXina. The authors gratefully acknowledge the LuxProvide teams for their expert support.

\bibliographystyle{IEEEtran}
\bibliography{bibliography}
\balance
\end{document}